\documentclass[twocolumn,showpacs,pra]{revtex4-1}
\usepackage{graphicx}
\usepackage{amsmath}
\usepackage{amssymb}
\begin{document}

\title{Engineering mesoscopic superpositions of superfluid flow}
\author{D. W. Hallwood and J. Brand}
\affiliation{Centre for Theoretical Chemistry, and Physics and New Zealand Institute for Advanced Study, Massey University, Private Bag 102904, North Shore, Auckland 0745, New Zealand.}

%=====================================================================
\begin{abstract}
Modeling strongly correlated atoms demonstrates the possibility to prepare quantum superpositions that are robust against experimental imperfections and temperature. Such superpositions of vortex states are formed by adiabatic manipulation of interacting ultracold atoms confined to a one-dimensional ring trapping potential when stirred by a barrier. Here, we discuss the influence of non-ideal experimental procedures and finite temperature. Adiabaticity conditions for changing the stirring rate reveal that superpositions of many atoms are most easily accessed in the strongly-interacting, Tonks-Girardeau, regime, which is also the most robust at finite temperature. NOON-type superpositions of weakly interacting atoms are most easily created by adiabatically decreasing the interaction strength by means of a Feshbach resonance. The quantum dynamics of small numbers of particles is simulated and the size of the superpositions is calculated based on their ability to make precision measurements. Experimental creation of strongly correlated and NOON-type superpositions with about 100 atoms seems feasible in the near future.
\end{abstract}
%=====================================================================

\maketitle

%=====================================================================
\section{Introduction}
%Research question:
Experimental advances in controlling ultracold atoms have led to leaps in the ability to control and create new types of quantum matter~\cite{greiner02,paredes04,hadzibabic06,lin09,ryu07}. A current challenge is to create larger superpositions for quantum technologies and for probing the foundations of quantum mechanics~\cite{arndt99,wal00,friedman00}. However, problems arise when attempting to make large superpositions, which improvement in experimental precision cannot overcome alone. 
% Conclusion
Here, we exploit the quantum properties of strongly correlated atoms to reduce the experimental precision required. The aim is to provide a path for creation of the largest superpositions with ultra-cold atoms so far.

%Motivation
Developing quantum technologies, such as precision measurement devices, constitutes an important goal in quantum physics research~\cite{giovannetti04}. Quantum entangled wave sources improve measurement accuracy without consuming more resources. Improved accuracy allows us to apply greater scrutiny to current fundamental theories and has practical applications, such as gyroscopes used for navigation in airplanes and satellites. Current devices predominantly use unentangled particles, due to their robustness against decoherence. However, some improvements have been demonstrated in gravitational-wave detectors using squeezed light~\cite{goda08}, and in atom interferometry with a Bose-Einstein condensate~\cite{esteve08}. If large robust quantum superpositions become available they will dramatically improve measurement precision~\cite{donner09,escher11,cooper10,hallwood09}. At a fundamental level, a microscopic understanding of the emergence of classical behavior in large objects is still needed. One route to improve our understanding is by investigating large superpositions~\cite{leggett85}.

%Literature
Currently the largest superpositions are found in superconducting loops split by one or more Josephson junctions, which are called flux qubits~\cite{wal00,friedman00}. They produce superpositions of different flux states penetrating the loop, or, equivalently, opposite currents flowing around the loop. The system was initially proposed by Leggett~\cite{leggett02} and much work has been done to understand how such large superpositions are possible~\cite{marquardt08,korsenbakken07,korsenbakken09}. Until recently, modeling of these types of systems relied on a phenomenological approach. However, by modeling an analogous system of ultracold atoms, previous work was able to show that strong interactions between particles allow a large coupling between the two current states~\cite{hallwood10}. This makes the superpositions far more robust then other many-particle superpositions, such as NOON states~\cite{gao10,nagata07,sackett00,hallwood07,franzer07}. 

%Results+validity
In this paper we study several experimental factors that affect the creation of a binary superposition of vortex states. Initially, we study how the superposition can be made. According to the proposal in Ref.~\cite{hallwood10}, experimental parameters must be evolved adiabatically after preparing the system in a unique ground state, such that the system is not excited. Two parameters are considered here: the stirring rate of the barrier and the interaction strength between the atoms. The system is most sensitive to the stirring rate of the barrier and we show how this can be increased to create the superposition. The fastest allowed route to formation of the superposition is found in the strongly interacting regime. Superpositions of strongly correlated states are, thus, more easily accessed than NOON states, which correspond to weak interactions. However, it seems possible to reach NOON states by a different route, which is to transform the strongly-correlated superposition to a NOON state by adiabatically reducing the interaction strength. By considering the adiabaticity conditions we find that this second approach is much more feasible. Two other factors are considered: how precisely the stirring rate must be tuned to obtain a balanced superposition, and how low the temperature of the system must be kept for the superposition to survive. Again, we see that the strongly-correlated state carries the greatest potential for the experimental realization of a superposition. Very recently the non-adiabatic creation of large angular momentum superpositions in the same system was suggested in Ref.~\cite{schenke11}.
%=====================================================================

%-------------------------------------------------------------------------------------------------------------------------
\begin{figure}[t]
\includegraphics[width=5.5cm]{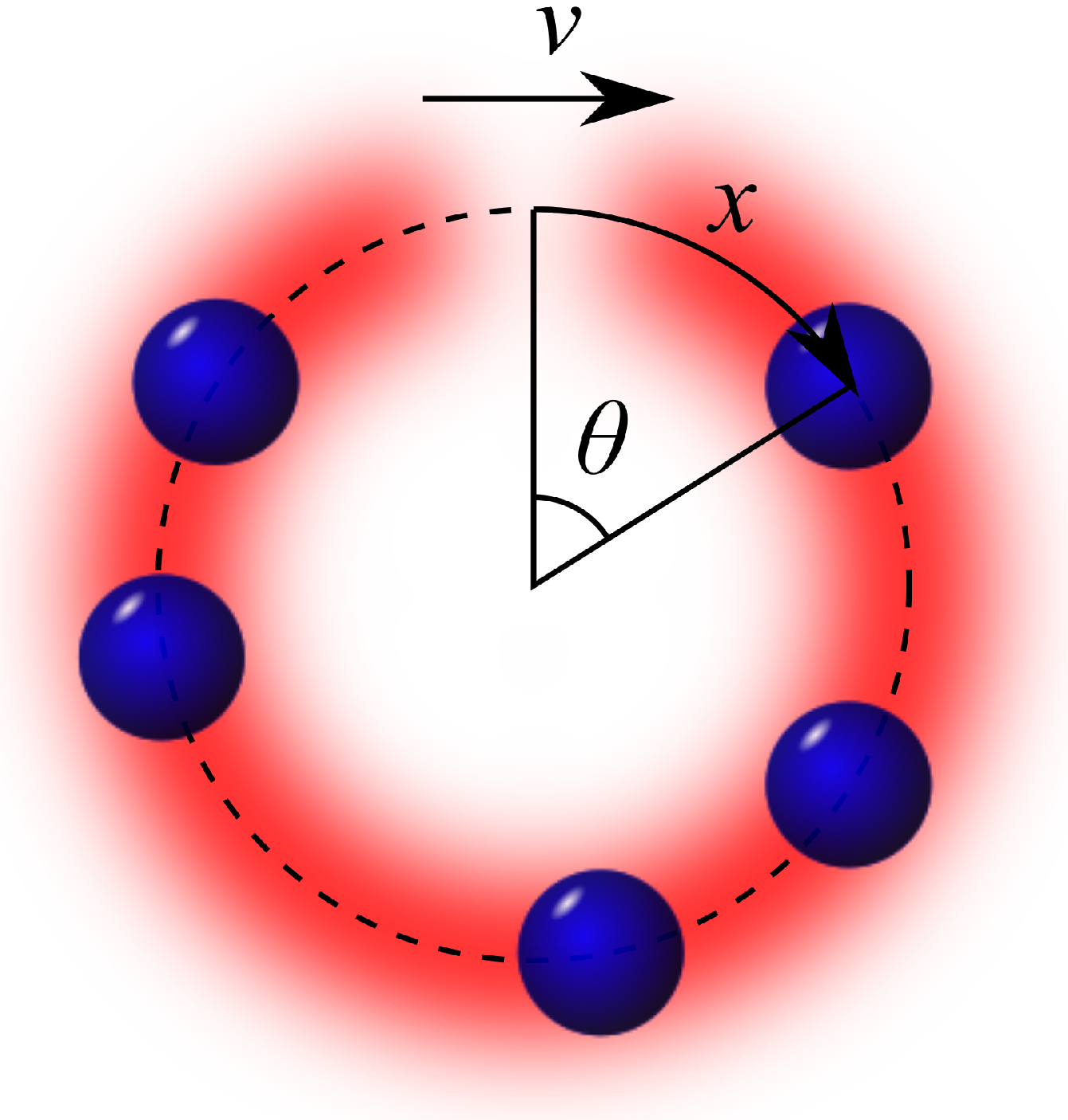}
\caption{ Ultra-cold atoms confined to a loop of circumference $L$. The position on the circumference of the loop is given by $x=L\theta$. The barrier is shown by the break in the ring and modeled by a $\delta$-function that moves with a tangential velocity of $v=\hbar \Omega/ML$. }
\label{fig:system}
\end{figure}
%--------------------------------------------------------------------------------------------------------------------------

%======================================================================
\section{The system}
A Hamiltonian describing $N$ interacting bosonic atoms confined to a toroidal trapping potential of circumference $L$ and stirred by a barrier can be written as
\begin{eqnarray}
H &=& \sum_{k=-\infty}^{\infty} E_0 \left( k-\frac{\Omega}{2\pi} \right)^2 \hat{a}^{\dag}_k \hat{a}_k + \frac{b}{L} \sum_{k_1,k_2=-\infty}^{\infty} \hat{a}^{\dag}_{k_1} \hat{a}_{k_2} \nonumber\\
&&+ \frac{g}{2L} \sum_{k_1,k_2,q=-\infty}^{\infty}  \hat{a}^{\dag}_{k_1} \hat{a}^{\dag}_{k_2}  \hat{a}_{k_1-q} \hat{a}_{k_2+q} \nonumber \\
&=& H_K + H_B + H_I.
\label{eq:H3}
\end{eqnarray} 
The three terms describe the kinetic energy of the atoms, $H_K$, the interaction between a barrier and the atoms, $H_B$, and the atom-atom interactions, $H_I$, respectively.  Here $\hat{a}^{\dag}_k$ ($\hat{a}_k$) creates (destroys) an atom with angular momentum $k\hbar$. The Hamiltonian is formulated in the co-rotating frame of the barrier, which is assumed to be narrow, and is described by a $\delta$-function with strength $b$. The tangential velocity of the barrier is $v=\hbar \Omega/ML$ along the circumference of the ring, where $\Omega$ is a dimensionless rotation rate. The smallest nonzero kinetic energy of a single particle is $E_0=2\pi^2\hbar^2/(ML^2)$, which provides a natural energy unit. The radial confinement is assumed to be tight and has a trapping frequency of $\omega_{\perp}$. The system can, therefore, be approximately described by a one-dimensional ring with coupling constant (or interaction strength)
\begin{equation}
g = \frac{2 \hbar^2}{M}\frac{a}{a_{\perp}} \left( a_{\perp} - \mathcal{C}a \right)^{-1},\;\;\;\;\; \mathcal{C} = 1.4603...,
\end{equation}
where $a$ is the s-wave scattering length and $a_{\perp} = \sqrt{\hbar/M\omega_{\perp}}$ characterizes the transverse confinement~\cite{olshanii_98}. The interaction strength can, therefore, be changed in two ways: the s-wave scattering length  can be tuned using a Feshbach resonance~\cite{inouye98, chin10}, or $a_{\perp}$ can be changed, which may lead to a confinement induced resonance~\cite{haller10}. This setup is schematically represented in Fig.~\ref{fig:system}. 

All numerical simulations are performed for five atoms, $N=5$, in a truncated Hilbert space with 18 angular momentum modes, where the interaction strength, $g$, is rescaled to account for the basis set truncation error~\cite{hallwood10, ernst11}. 

A general wavefunction can be written as a superposition of terms with different total angular momentum,
\begin{equation}
|\Psi\rangle = \sum_{K=-\infty}^{\infty} C_K|K\rangle
\label{eq:tot_mom_state}
\end{equation}
where 
\begin{equation}
|K\rangle=\sum_{\vec{n}}{\LARGE}^{(K)}A_{\vec{n}}|\vec{n}\rangle
\label{eq:mom_state}
\end{equation}
is the sum over the multi-index $\vec{n}=(...,n_{-1},n_{0},n_1,...)$ with fixed particle number, $\sum_k n_k = N$, and $\sum^{(K)}$ implies the additional constraint $\sum_k n_k k = K$ fixing the total angular momentum. 
Using the creation operators $\hat{a}_k^{\dag}$ we can construct the permanents,
\begin{equation}
|\vec{n}_K\rangle = \prod_k \frac{1}{\sqrt{n_k!}} \left( \hat{a}_k^{\dag} \right)^{n_k}|{\rm vac}\rangle.
\end{equation}
Therefore, the probability that the state~(\ref{eq:tot_mom_state}) has total angular momentum $K\hbar$ is $|C_K|^2$. In the following, $|K\rangle$, which is labelled by a single index referring to the total angular momentum, is a superposition of permanents with total angular momentum $K$, while $|...,n_{-1},n_0,n_1,...\rangle_{...,-1,0,1,...}$, which is labelled by several indices referring to the occupation of the angular momentum modes, will describe a permanent, where $n_k$ atoms occupy each mode $k$.

We wish to make large superpositions. One example is the NOON state, which is a superposition of all $N$ particles being in one mode or all $N$ particles being in another mode. The NOON state can be written as
\begin{equation}
|\Psi\rangle_{\pm} =  \frac{1}{\sqrt{2}}\left(|N,0\rangle_{0,1} \pm |0,N\rangle_{0,1} \right),
\label{eq:noonSP}
\end{equation}
where each term in the ket represents the occupation in a particular mode. The terms here represent the number of atoms in the 0 and $\hbar$ angular momentum modes. Schemes to create NOON states using ultra-cold atoms have been proposed in a double well~\cite{dunningham01,franzer07}, a ring lattice of three sites~\cite{dunningham06,hallwood06} and a ring lattice of four sites~\cite{nunnenkamp08}. 

Here we consider a broader range of many-particle superpositions, which are binary superpositions of eigenstates of the total angular momentum operator, as opposed to the two mode NOON state. This is more in keeping with Schr\"{o}dinger's idea of a cat being in a superposition of alive and dead. In Ref.~\cite{hallwood10} it was demonstrated that in systems described by Eq.~(\ref{eq:H3}) it is possible to make superpositions of total angular momentum of the form,
\begin{equation}
|\Phi\rangle_{\pm} = \frac{1}{\sqrt{2}}\left(|0\rangle \pm |N\rangle \right)
\label{eq:K_SP}
\end{equation}
by applying a rotation rate of $\Omega=\pi$ and having an interaction strength above a certain value. Here $N$ is the number of atoms in the system. Therefore, the NOON state is just a special case of this superposition. In this paper we study the experimental control required to make the superposition in Eq.~(\ref{eq:K_SP}), but first the spectrum of possible states is briefly described.
%==================================================================

%--------------------------------------------------------------------------------------------------------------------
\begin{figure}[t]
\includegraphics[width=8cm]{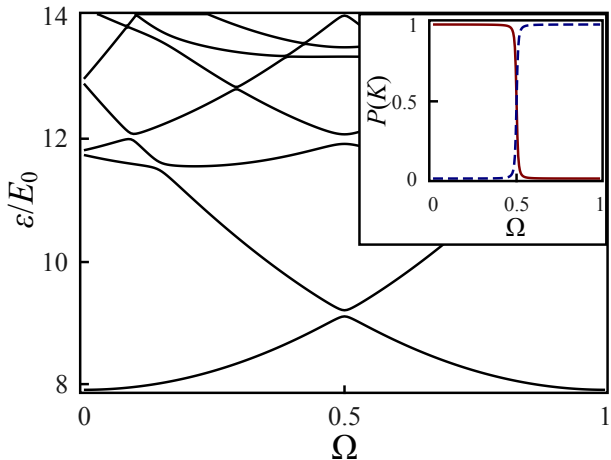}
\caption{  The energy spectrum of five strongly correlated atoms confined to a ring trapping potential and large barrier as a function of $\Omega$ (subsequent results use a smaller barrier height of $b/L=0.008E_0$ for increased numerical accuracy).  At $\Omega = \pi$ there is an avoided crossing in the lowest two levels and this is the point where the superposition $|0\rangle +|N\rangle$ is found. The inset shows the probability that the ground state has total angular momentum $K=0$, $P(0)$, as a full line, and total angular momentum $K=N\hbar$, $P(N)$, as a dashed line. At $\Omega = \pi$ the probability of the two states is 0.5.}
\label{fig:EvOm_PKvOm}
\end{figure}
%---------------------------------------------------------------------------------------------------------------------

%===================================================================
\section{Multi-particle superpositions}

% Non-interacting atoms
\emph{Non-interacting atoms -} For simplicity, consider just the 0 and $\hbar$ angular momentum modes, which is a valid approximation for a small barrier height. The three terms in the Hamiltonian~(\ref{eq:H3}) play three different, but crucial roles in creating the superpositions. The kinetic energy term, $H_K$, changes the energy of the two modes as the stirring rate of the barrier changes. At $\Omega=\pi$ the two modes become degenerate and it is at this value that a superposition can form. For $N$ non-interacting atoms with a finite barrier height a superposition of the form
\begin{eqnarray}
|\phi\rangle	&=& \frac{1}{\sqrt{2^N N!}}\left( \hat{a}_0^{\dag} + \hat{a}_1^{\dag}   \right)^N |\textrm{vac} \rangle \nonumber\\
			&=& \frac{1}{\sqrt{2^N}} \sum_{r=0}^N\sqrt{\frac{N!}{(N-r)!r!}} |N-r,r\rangle_{0,1}
\label{eq:binSP}
\end{eqnarray}
is created. This is just a product of single particle superpositions and not the state we are interested in here. Exact results for non-interacting atoms beyond the two mode approximation were derived in Ref.~\cite{hallwood10}.

Note the importance of the barrier term in Eq.~(\ref{eq:H3}), $H_B$. This is the only part of the Hamiltonian that couples states with different total angular momentum. Without a barrier, the ground state is doubly degenerate at $\Omega = \pi$. With a barrier, the degeneracy is lifted leaving an energy gap between the ground and first excited state. 

% NOON state
\emph{NOON state -} Equation~(\ref{eq:binSP}) is a superposition of states of the form $|N-r,r\rangle_{0,1}$, because all these states are degenerate for $g=0$. The degeneracy is lifted by the interaction term, ${}_{0,1}\langle N-r,r|H_I|N-r,r\rangle_{0,1}=\frac{g}{2L}\left[ N(N-1) + 2r(N-r) \right]$, leaving $|N,0\rangle$ and $|0,N\rangle$ as the lowest degenerate energy states. Experimentally, the interaction strength can be changed using a Feshbach resonance~\cite{inouye98}. Again, $H_B$ couples the $|N,0\rangle$ and $|0,N\rangle$ states and the ground and first excited state become the superpositions in Eq.~(\ref{eq:noonSP}) for a large enough interaction strength.

% Tonks-Girardeau
\emph{Tonk-Girardeau superposition -} For stronger interactions, the coupling to momentum modes other than 0 and $\hbar$ becomes significant and we can no longer write the state as in Eq.~(\ref{eq:noonSP}). However, the ground and first excited states remain in superpositions of total angular momentum as described by  Eq.~(\ref{eq:K_SP}). In the regime of strongly interacting bosons at low densities, and a tightly confining wave guide, there is a one-to-one correspondence with spinless non-interacting fermions. The bosonic atoms are said to have undergone fermionisation, because they can no longer pass one another. This system is call a Tonks-Girardeau gas~\cite{girardeau60}.  The energy spectrum is the same as that for spinless non-interacting fermions. However, although atoms cannot be in the same position, they can have the same momentum, so the single-particle momentum distribution is different to that of spinless fermions. Away from zero momentum, the distribution spreads as $\propto 1/\sqrt{|k|}$ over an infinite range even for zero temperature~\cite{Lenard}.

The energy spectrum of the Tonks-Girardeau gas is calculated using the single particle spectrum in the same way as for spinless fermions. Consider an odd number of atoms (the calculation is more complicated for even numbers, but still possible~\cite{lieb63}), where the single-particle energies are given by $\varepsilon_n$, $\varepsilon_0$ is the lowest energy, and the energies are in ascending order. Therefore, the ground state energy is then $\sum_{n=0}^{N-1}\varepsilon_n$, the first excited state is $\sum_{n=0}^{N-2}\varepsilon_n+\varepsilon_N$, and the energy difference is $\Delta E = \varepsilon_N-\varepsilon_{N-1}$. Even in the Tonks-Girardeau regime the binary superposition of total angular momentum is still created and Ref.~\cite{hallwood10} demonstrated that this superposition is most robust against the loss of atoms than NOON states.

% Energy gap
\emph{Energy gap -} The energy spectrum of the system provides a useful insight into the experimental control required to create these superpositions. Figure~\ref{fig:EvOm_PKvOm} demonstrates how rotation of the barrier effects the state of the system. At the rotation rate that produces $\Omega=\pi$, an avoided crossing is formed between the ground and first excited state, which is a signature of a superposition. This can be understood by considering a simple two state model with states $|0\rangle$ and $|N\rangle$, which will be useful later. Writing the wavefunction $|\psi\rangle = \alpha|0\rangle + \beta|N\rangle$ in matrix form, 
\begin{eqnarray}
|\psi\rangle = 
\left(
\begin{array}{c}
 \alpha  \\
 \beta
\end{array}
\right),
\end{eqnarray}
the effective Hamiltonian, derived from the Hamiltonian in Eq.~(\ref{eq:H3}), is,
\begin{eqnarray}
H = 
\left(
\begin{array}{cc}
 0 & \Delta  \\
 \Delta  & E_0 N \left( 1- \Omega/\pi \right),
\end{array}
\right)
\label{eq:Heff}
\end{eqnarray}
where $\Delta$ is the coupling between the two states and a constant energy term has been ignored. The eigenstates of the Hamiltonian are,
\begin{eqnarray}
|\psi_0\rangle = 
\left(
\begin{array}{c}
 C_a (\Omega)  \\
 C_b (\Omega)
\end{array}
\right), \;\;\;\;\;
{\rm and} \;\;\;\;\;
|\psi_1\rangle = 
\left(
\begin{array}{c}
 C_b (\Omega)  \\
 -C_a (\Omega)
\end{array}
\right),
\end{eqnarray}
where
\begin{eqnarray}
 |C_{a,b}|^2 = \frac{1}{2} \pm \frac{E_0 N (\Omega -\pi)}{2\sqrt{\Delta^2 + E_0^2N^2(\Omega -\pi)^2}},
 \label{eq:coeffs}
\end{eqnarray}
and the energy difference between the two energy levels is $\hbar \omega_{01} =  \sqrt{4\Delta^2+E_0^2N^2(1-\Omega/\pi)^2}$. Note that at the avoided crossing, $\Omega = \pi$, the energy gap is just twice the coupling between the two states, $\Delta E = 2\Delta$. The subplot in Fig.~\ref{fig:EvOm_PKvOm} shows the probability of finding the system in the $|0\rangle$ and $|N\rangle$ states. Notice how quickly the superposition becomes unbalanced when $\Omega$ is detuned from $\pi$. The precision of tuning $\Omega$ required to create the superposition is discussed in detail in Sec.~\ref{sec:pot_dec}. 

%---------------------------------------------------------------------------------------------------------
\begin{figure}[t]
\includegraphics[width=8cm]{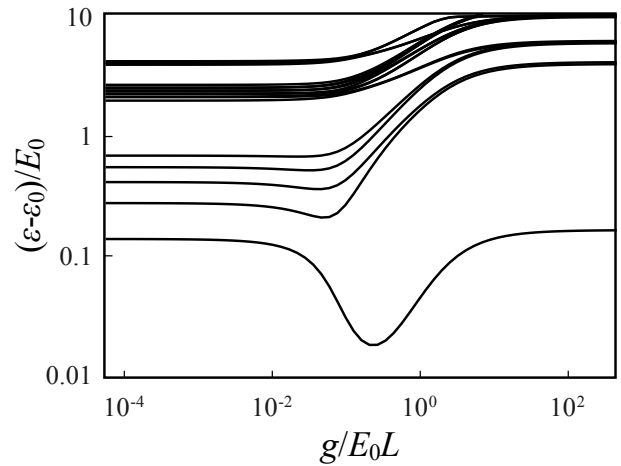}
\caption{ The energy spectrum of atoms confined to a ring trapping potential as a function of interaction strength $g$ with $b/L=0.08 E_0$ and $\Omega=\pi$. Energies are given relative to the ground-state energy. }
\label{fig:Evg}
\end{figure}
%----------------------------------------------------------------------------------------------------------

% Dependence on the interaction strength
\emph{Dependence on the interaction strength -} Figure~\ref{fig:Evg} shows how atom-atom interactions effect the energy spectrum, where all energies are taken relative to the ground state. Notice the dip in the lowest line which signals a near degeneracy of the ground and first excited state. We find that this is the interaction strength required to make the NOON state. The small energy gap shows that the coupling between the $|N,0\rangle_{0,1}$ and $|0,N\rangle_{0,1}$ states is very weak. The energy spectrum, together with the temperature determines the thermal state of the system, which is discussed in Sec.~\ref{sec:therm}. Another factor that is strongly dependent on the energy gap are the conditions to reach adiabaticity upon changing parameters. Here we analyze how $\Omega$ can be evolved in time to create the superposition. This method turns out to only be feasible for creating superpositions in the strongly interacting regime. We study a more realistic method for making a NOON state that starts with a superposition in the strongly interacting regime, and then slowly reduces the interaction strength to that needed to make a NOON state.

%------------------------------------------------------------------------------------------------------------
\begin{figure}[t]
\includegraphics[width=8cm]{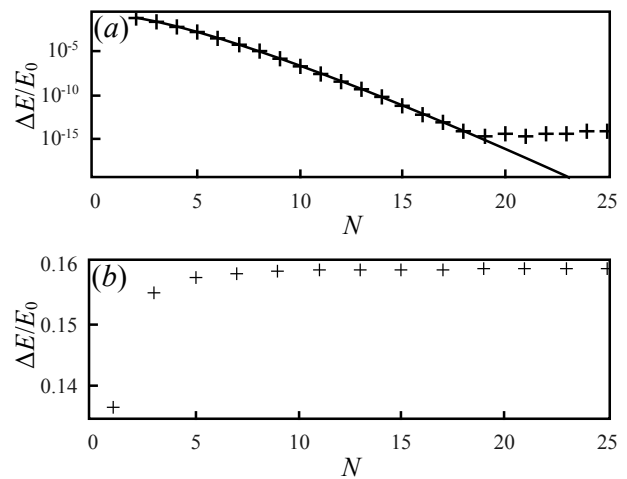}
\caption{  The energy level splitting at the avoided crossing, $\Delta E$, is plotted against $N$ for both ($a$) the NOON state (also found in Ref.~\cite{hallwood10}) and ($b$) the Tonks-Girardeau state. Part ($a$) shows analytic result for two modes (solid line) and the numerical result for two modes ($+$), where $b/L = 0.008E_0$ and $g/L = 4\pi b\sqrt{N} /L(N-1) $. The numerical result breaks down for $N > 18$ due to limited numerical accuracy. The analytic result is still valid beyond this point. Part ($b$) bottom figure shows the analytic result for $b/L=0.08E_0$. For large $N$ and a small barrier, $\Delta E$ asymptotes towards $\Delta E = 2b/L$.}
\label{fig:DEvN}
\end{figure}
%-------------------------------------------------------------------------------------------------------------

% Dependence on the number of atoms
\emph{Dependence on the number of atoms -} To create large superpositions we must also consider how the system changes as the number of atoms is increased. Figure~\ref{fig:DEvN} shows the energy gap between the ground and first excited states as a function of the number of atoms. These results were obtained analytically as described in Ref.~\cite{hallwood10}. The NOON state energy is calculated from a two mode system with the 0 and $\hbar$ angular momentum modes, which gives,
\begin{eqnarray}
\Delta E = \frac{b^N}{g^{N-1}}\frac{2}{L}\frac{N}{(N-1)!}
\end{eqnarray}
where we require 
\begin{eqnarray}
\frac{b\sqrt{N}}{L} \ll \frac{gN}{2L} \ll E_0
\label{eq:condition}
\end{eqnarray}
for the NOON state to be created. The energy gap, and therefore the coupling between the two states, rapidly decreases as the number of atoms is increased as seen in Fig.~\ref{fig:DEvN} ($a$). This makes the superposition very fragile. 

The energy gap in the Tonks-Girardeau regime is calculated using the Bose-Fermi mapping. Therefore, only the single particle energy spectrum is needed. At $\Omega = \pi$ this is given by the roots of
\begin{eqnarray}
\frac{2\pi \hbar^2 \alpha_{\mu}}{MLb} = -\tan(\pi\alpha_{\mu})
\end{eqnarray}
for odd $\mu$ and $\alpha_{\mu} = (\mu + 1)/2$ for even $\mu$, where the energy is just $\varepsilon_{\mu} = \alpha_{\mu}E_0$~\cite{hallwood10}. We find that the energy gap actually increases with atom number to $2b/L$ for a small barrier height, $b/L \ll NE_0$ as shown in Fig.~\ref{fig:DEvN} ($b$), and reaches a maximum of $ (N-1/2)E_0/2$ for a large impenetrable barrier, $b/L \gtrsim NE_0$. The scaling of $\Delta E$ with particle number is, thus, much more favorable in the Tonks-Girardeau regime than for the NOON state. 

% Other factors
\emph{Particle loss -} Particle loss has also been shown to degrade multi-particle superpositions~\cite{hallwood10,cooper11}. NOON states are destroyed with the loss of a single particle if the state of the particle is known. However, for strongly interacting atoms the effect is less detrimental. This is because the single particle momentum distribution for the $|0\rangle$ and $|N\rangle$ states have a large overlap. Consequently, the knowledge of the angular momentum of a single atom is not enough to determine the constituent of the superposition where the atom came from. The loss of a single atom with known angular momentum, thus, only degrades the superposition, but does not destroy it completely.

\emph{Barrier shape -} Another complicating factor when considering the experimental feasibility of the setup is a finite width of the barrier. A barrier of finite width was considered in Ref.~\cite{nunnenkamp10}, where it was found that the energy gap decreases exponentially with the number of atoms. This can be understood simply by replacing the $\delta$-function barrier by the Gaussian
\begin{eqnarray}
V(x) = \frac{b}{\sqrt{2\pi}\sigma} \exp\left(- \frac{x^2}{2\sigma^2} \right),
\label{eq:gaussian}
\end{eqnarray}
where $\sigma$ defines the width of the Gaussian and $b$ is the area of the potential.  To first order in $b$ and for $\sigma \ll L$, the energy gap is given by
\begin{eqnarray}
\Delta E = 2\frac{b}{L} \exp\left(- 2\pi^2 \frac{\sigma^2}{L^2}N^2 \right).
\label{eq:DENOON}
\end{eqnarray}
which was found previously for a related system in Ref.~\cite{nunnenkamp10}.
If we want the reduction in the energy gap, due to a finite barrier width, to be  less than $1/e$ we require
\begin{eqnarray}
\sigma < \frac{L}{\sqrt{2}\pi N}.
\label{eq:width}
\end{eqnarray}
In contrast, a $\delta$-function barrier couples all angular momentum modes equally. 

In order to experimentally achieve efficient coupling the barrier width needs to be decreased with increasing particle number. Another solution could be to design a potential that can efficiently couple the relevant momentum modes for a given number of atoms, e.g. by Bragg reflection off an optical ring lattice with commensurate filling of atoms~\cite{hallwood07} or laser speckle. It should be noted that Ref.~\cite{nunnenkamp10} considers a different system, with atoms confined to a ring lattice at low filling rather than a continuous loop. Comparison between the results of Ref.~\cite{nunnenkamp10}, on the one hand, and those of Ref.~\cite{hallwood10} and this paper, on the other hand, suggest that the continuous loop is a preferable system for experimentally creating large superpositions.
%================================================================

%================================================================
\section{Adiabatic evolution}	%NEW -> CAUX\\
To create the superposition of total angular momentum the experimental parameters must be evolved slowly enough so the system has only a small probability of being excited out of the ground state. However, the length of time the system can be evolved for is limited by the coherence time of the atoms. Therefore, a compromise must be made that allows the system to be evolved sufficiently slowly so the probability of an excitation is small, and fast enough so the coherence of the system is not lost. Here we study the evolution rate necessary to achieve this for $\Omega$ and $g$.

In general, if we evolve a quantity $\zeta$ at a linear rate, $d\zeta/dt = \mbox{constant}$, for a time interval $[ t_0,t_f ]$, then we can approximate the probability that the system is excited to state $j$ by~\cite{dagnino09, messiah99}
\begin{eqnarray}
P_{0\rightarrow j} \le \max_{\zeta \in [\zeta_0,\zeta_f]} \left( \frac{\alpha_{0j}}{\omega_{0j}} \right)^2=\varepsilon^2,
\label{eq:P01}
\end{eqnarray}
where $\hbar\omega_{0j} =E_j-E_0$,
\begin{equation}
\alpha_{0j} = \langle \Psi_j | \frac{d|\Psi_0\rangle}{dt}=\frac{d\zeta}{dt}F(\zeta)
\label{eq:alpha01}
\end{equation}
$F(\zeta) = \langle \Psi_j| \frac{d|\Psi_0\rangle}{d\zeta}$, and $\zeta_0$ and $\zeta_f$ are the initial and final values of $\zeta$, respectively. We choose the value of $\zeta$ that maximizes the ratio $\alpha_{0j}/\omega_{0j}$ in order to obtain an upper bound on the transition probability. The probability that the system is not excited to state $j$ is, therefore, approximately $1-\varepsilon^2$. In what follows, the excitation to the first excited state is by far the most probable, so higher excited states can be ignored.

%= = = = = = = = = = = = = = = = = = = = = = = = = = = = = = = = = = = = = = = = = = = = = = = = =
\subsection{Creating the superposition}
\label{sec:SPcreate}
To create the superposition, $\Omega$ must be changed from 0 to $\pi$. For an interaction strength equal to, or greater than that required to make a NOON state, the superposition in Eq.~(\ref{eq:K_SP}) is created and this is what we consider here. The maximum probability for a transition out of the ground state is found at $\Omega = \pi$ and this is to the first excited state. Therefore we only need to consider the states $|0\rangle$and $|N\rangle$. The effective Hamiltonian is given by Eq.~(\ref{eq:Heff}). Here the coupling is calculated numerically using the relation $\Delta = \Delta E/2$. From Eq.~(\ref{eq:coeffs}) we find
\begin{eqnarray}
F(\pi) =  \frac{N}{2\pi^2}\frac{E_0}{\Delta E}.
\label{eq:Fn}
\end{eqnarray}
Therefore, to evolve $\Omega$ with only a small probability of exciting the system out of the ground state we need
\begin{eqnarray}
P_{0\rightarrow 1} \le \frac{d\Omega}{dt} \frac{N\hbar}{2\pi^2}\frac{E_0}{\Delta E^2}  \ll 1.
\end{eqnarray}
This shows that the system can be evolved quicker for smaller numbers of atoms and a larger energy gap at the avoided crossing. For the Tonks-Girardeau state with a large number of atoms and a small barrier this is
\begin{eqnarray}
\frac{d\Omega}{dt}  \ll  \frac{b^2}{E_0\hbar L^2}\frac{8\pi^2}{N},
\end{eqnarray}
while for a large barrier this is
\begin{eqnarray}
\frac{d\Omega}{dt}  \ll  \frac{\pi^2 N E_0}{\hbar}.
\label{eq:domdg1}
\end{eqnarray}
For the NOON state we find,
\begin{eqnarray}
\frac{d\Omega}{dt}  \ll  \frac{4\pi}{\hbar E_0 L^2}\frac{b^{2N}}{g^{2(N-1)}}\left(\frac{e}{N}\right)^{2N}
\label{eq:omrateNOON}
\end{eqnarray}
where Eqs.~(\ref{eq:DENOON}) and the Stirling's approximation has been used, $N! \approx \sqrt{2\pi N}(N/e)^N$. As we can see, the scaling for the NOON state is far worse than the Tonks-Girardeau state.
%= = = = = = = = = = = = = = = = = = = = = = = = = = = = = = = = = = = = = = = =

%= = = = = = = = = = = = = = = = = = = = = = = = = = = = = = = = = = = = = = = =
\subsection{Creating a NOON state}
We have shown that adiabatically changing $\Omega$ to create the NOON state is only possible for small numbers of atoms, because the near degeneracy at $\Omega = \pi$ makes the evolution time much longer than an experimentally realizable coherence time. Another possibility is to create a superposition in the Tonks-Girardeau regime first and then adiabatically decrease the interaction strength to that required for the NOON state. Another advantage of this scheme is that the barrier can be removed, which eliminates a potential source of decoherence due to coupling between states with different total angular momentum. This must be done at a rate that does not change the superposition significantly, while slow enough not to excite the system. Now there is no longer coupling between states with different angular momentum. 

The properties of the ground state change markedly when the interaction strength is reduced from the Tonks-Girardeau regime to the interaction strength needed for the NOON state. However, we can get an idea of how long it will take to create the NOON state by looking at the most sensitive part of the evolution, which is at the NOON state. Because there is no coupling to states with different total angular momentum we will just consider the states with zero total angular momentum. The two most important states are thus $|0,N,0\rangle_{-1,0,1}$ and $|1,N-2,1\rangle_{-1,0,1}$, and we assume a linear ramping in time of the interaction strength $g$. 

Again, writing the wavefunction $|\psi\rangle = \alpha|0,N,0\rangle_{-1,0,1} + \beta|1,N-2,1\rangle_{-1,0,1}$ in matrix form
\begin{eqnarray}
|\psi\rangle = 
\left(
\begin{array}{c}
 \alpha  \\
 \beta
\end{array}
\right),
\end{eqnarray}
the effective Hamiltonian, derived from the Hamiltonian in Eq.~(\ref{eq:H3}), is
\begin{eqnarray}
H = 
\left(
\begin{array}{cc}
 0 & g/2L  \\
 g/2L  &  2E_0 +  \frac{g}{2L}(4N-6)
\end{array}
\right).
\label{eq:Heff5}
\end{eqnarray}
By calculating the eigenenergies and states of the Hamiltonian we come to the solution
\begin{eqnarray}
F(g)  =\frac{E_0}{L} \frac{\sqrt{N(N-1)}}{\hbar^2 \omega_{01}^2},
\end{eqnarray}
where
\begin{eqnarray}
\hbar \omega_{01} = \sqrt{\left[2E_0 \!+\! \frac{g}{2L}(4N\!-\!6)\right]^2 \!+\! 4\left(\frac{g}{2L}\right)^2N(N\!-\!1)}
\end{eqnarray}
is the energy difference of the two eigenstates of the Hamiltonian. 

The probability of a transition is maximised by taking $g$ to zero, however, we only need to reach the interaction strength that creates the NOON state. To make the probability of exciting the system small we must satisfy the condition
\begin{eqnarray}
\frac{dg}{dt} \ll \frac{L}{E_0} \frac{\hbar^2\omega_{01}^3}{ \sqrt{N(N-1)}}.
\label{eq:dgdt}
\end{eqnarray}
For large $N$ this is
\begin{eqnarray}
\frac{dg}{dt} \ll \frac{5\sqrt{5}}{\hbar E_0} \frac{g^3}{ L^2} N^2.
\label{eq:dgdt}
\end{eqnarray}
This scaling is far better than the scaling found in Eq.~(\ref{eq:omrateNOON}), so this method represents a substantially more realistic method for producing large NOON states.
%============================================

%============================================
\section{Size of the superposition}
There have been several approaches to define the size of a superposition~\cite{leggett_87,korsenbakken07,marquardt08}. Broadly speaking these fit into two categories: a measure of the entanglement in the system and a measure of the physical difference of the two parts of the superposition. Here we take a different approach that is based on the ability of the system to make a precision measurement. Superposition states are useful in interferometry for estimating parameters. Larger superpositions are better and give more precision. This can be quantified using the quantum Fisher information (QFI), $F_Q$, which describes the maximum amount of information that can be extracted about a certain quantity and is the standard parameter used to determine the usefulness of a system for making a precision measurement~\cite{Rao1945,Cramer1946,Helstrom1976}. This is related to the uncertainty of the measurement via the Cram\'er-Rao bound, $\delta\phi \ge 1/\sqrt{F_Q}$~ \cite{Helstrom1976}. It has been shown that large superpositions can be used to perform precision measurements beyond the classical limit~\cite{goda08, esteve08}. Here the QFI is calculated to determine how the superposition is degraded under non-ideal experimental conditions. The influence of particle loss on the QFI was already considered in Ref.~\cite{cooper11}.

Consider a superposition of two states with different total angular momentum,
\begin{eqnarray}
|\Psi\rangle = \frac{1}{\sqrt{2}}\left( |K_1\rangle + |K_2\rangle \right)
\label{eq:bin_sup}
\end{eqnarray}
A sudden change in the rotation rate of the barrier will change the energy of the two states and they will acquire different phases,
\begin{eqnarray}
|\Psi(\phi)\rangle = \frac{1}{\sqrt{2}}\left(e^{-iK_1\phi} |K_1\rangle +e^{-iK_2\phi} |K_2\rangle \right)
\label{eq:bin_supPhase}
\end{eqnarray}
where a global phase factor due to the interaction and barrier terms have been ignored. The phase $\phi = \hbar \Omega t/ML^2 $ is the rotation angle of the barrier acquired during a time $t$ by time evolution with the Hamiltonian~(\ref{eq:H3}). From the $\phi$ dependence of the wavefunction, the QFI can be calculated, which measures how well the state can determine $\phi$.  The QFI is given by 
\begin{equation}
F_Q=\text{Tr}[\rho(\phi)A^2],
\label{eq:FQmix}
\end{equation}
where $\rho(\phi)$ is the density matrix of the system, and $A$ is the symmetric logarithmic negativity, defined as
\begin{equation}
\frac{\partial \rho(\phi)}{\partial \phi} = \frac{1}{2}[A\rho(\phi) + \rho(\phi)A].
\end{equation}
In the eigenbasis of $\rho(\phi)$ this is $(A)_{ij} = 2[\rho'(\phi)]_{ij}/(\lambda_i+\lambda_j)$, where $\lambda_{i,j}$ are the eigenvalues of $\rho(\phi)$ and $\rho'(\phi) = \partial \rho(\phi)/\partial \phi$.  If $\lambda_i+\lambda_j=0$ then $(A)_{ij}=0$.  For a pure state, $|\Psi(\phi)\rangle$, the QFI is
\begin{equation}
F_Q=4\left[ \langle\Psi'(\phi)|\Psi'(\phi)\rangle - \left|  \langle\Psi'(\phi)|\Psi(\phi)\rangle \right|^2 \right],
\label{FQpure}
\end{equation}
where $|\Psi'(\phi)\rangle = \partial|\Psi(\phi)\rangle/\partial \phi$~\cite{Braunstein1994}. 

For the superposition of total angular momentum states~(\ref{eq:bin_sup}), the quantum Fisher information is $(K_1-K_2)^2$. For the case of non-interacting atoms described in Eq.~(\ref{eq:binSP}) we do not have a binary superposition, however we can still calculate the QFI using Eq.~(\ref{FQpure}) and obtain $F_Q = N$. This is called the shot noise limit and is the maximum QFI achievable by unentangled particles or independent measurements. A significant increase in the QFI is achieved when using NOON states, such as the one given in Eq.~(\ref{eq:noonSP}). The QFI is now $F_Q = N^2$. This is called the Heisenberg limit and is the maximum precision predicted by quantum mechanics~\cite{Yurke86, Caves1981, Dowling1998, Lee2002}. Furthermore, all states that can be written in the form of Eq.~(\ref{eq:K_SP}) also have this QFI. This was first demonstrated in Ref.~\cite{cooper11}.

%= = = = = = = = = = = = = = = = = = = = = = = = = = = = = = = = = = = =
\subsection{Potential decoherence}
\label{sec:pot_dec}
A major difficulty in creating large superpositions is their extreme sensitivity to experimental inaccuracies. This was first investigated for NOON states created in a ring lattice~\cite{hallwood07} and subsequently for NOON states created in a double well, where this effect was aptly named \emph{potential decoherence}~\cite{franzer07}. In this section, the investigation is not limited to NOON states, but instead looks at the range of superpositions from uncorrelated to strongly interacting atoms. The investigation is restricted to imperfections in the rotation rate of the barrier, because the system is most sensitive to this parameter.

The same effective Hamiltonian described in Sec.~\ref{sec:SPcreate} and given by Eq.~(\ref{eq:Heff}) can be used here. Consider interaction strengths strong enough to create a NOON state and stronger. This ensures that the ground state is a binary superposition of total angular momentum when $\Omega = \pi$, so only the states $|0\rangle$ and $|N\rangle$ are relevant. To measure the effect of detuning of $\Omega$ away from $\pi$ we study how the QFI changes. 

\begin{figure}[t]
\includegraphics[width=8.5cm]{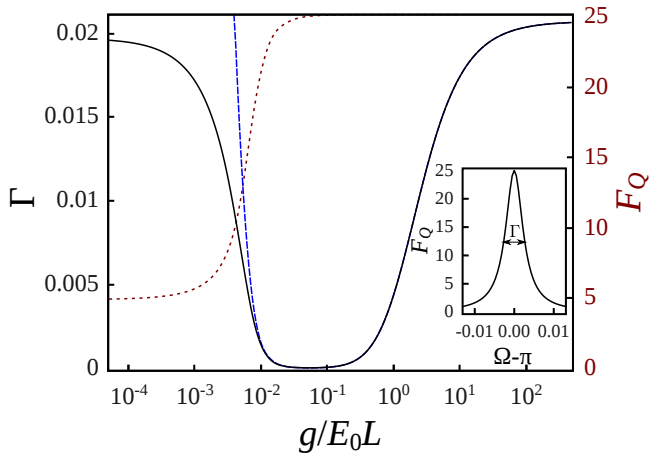}
\caption{  The inset shows the QFI as a function of $\Omega$ in the Tonks-Girardeau regime. Shown in the inset is the full width half maximum, $\Gamma$. In the main figure, $\Gamma$ is plotted on the left axis against the interaction strength, $g$, for for $b/L=0.08E_0$. The solid line shows the numerical result and the dashed line the semi-analytic result given by Eq.~(\ref{eq:gamma}). The right axis shows the QFI for different interaction strengths when $\Omega = \pi$ (dotted line). }
\label{fig:dP0dOmvg}
\end{figure}

Again, the coupling term is given by half the energy gap at $\Omega=\pi$, and can be obtained numerically. The eigenvalues and states of the Hamiltonian are given in Sec.~\ref{sec:SPcreate}, thus 
\begin{eqnarray}
F_Q =  \frac{(\Gamma/2)^2N^2}{(\Gamma/2)^2+(\Omega-\pi)^2}.
\label{eq:lorentzdist}
\end{eqnarray}
This is a function of $\Omega$ with the shape of a Lorentzian distribution~\cite{barnett97}, width
\begin{eqnarray}
\Gamma=\frac{\pi \Delta E}{E_0 N},
\label{eq:gamma}
\end{eqnarray}
and a maximum amplitude of $N^2$. Equation~(\ref{FQpure}) is simplified to $F_Q=4\sum_K\left[ K^2 |C_K|^2 (1-|C_K|^2)\right]$ for the two states considered here. We have calculated the QFI numerically and plotted it against the analytic result in Fig.~\ref{fig:dP0dOmvg}.  The numerical and analytic results give good agreement except for small interactions. This is because the two state approximation fails in this regime. For large interactions, $g$, the value of the QFI is $N^2$. However this drops to $N$ for small $g$ showing that the system is uncorrelated (see Fig.~\ref{fig:dP0dOmvg}). 

$\Gamma$ must be large to create states that are less sensitive to changes in $\Omega$. This is realised when the energy gap is large and the number of atoms is small. $\Delta E$ is approximately constant as a function of atom number for large numbers of atoms in the Tonks-Girardeau case, while it decreases rapidly for the NOON state (see Fig.~\ref{fig:DEvN}). The gradients of the energy levels as a function of $\Omega$ are proportional to $N$, which explains the number dependence in the denominator of Eq.~(\ref{eq:gamma}). This will present a limiting factor for experimentally realizing this state for large numbers of atoms.
%= = = = = = = = = = = = = = = = = = = = = = = = = = = = = = = = = = = = =

%-------------------------------------------------------------------------------------------------------------
%-------------------------------------------------------------------------------------------------------------

%-------------------------------------------------------------------------------------------------------------
\begin{figure}[t]
\includegraphics[width=8cm]{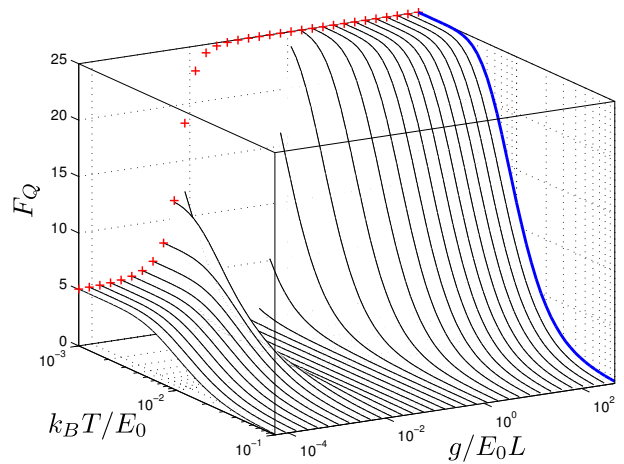}
\caption{ (Color online) QFI as a function of temperature and interaction strength. Each line represents a particular interaction strength. The thick blue line show the analytic result for a Tonks-Girardeau superposition of Eq.~(\ref{eq:QFI_TG}) and the red crosses show the QFI at $T=0$ from Fig.~\ref{fig:dP0dOmvg}. The barrier height is $b/L=0.08E_0$.}
\label{fig:QvT}
\end{figure}
%-------------------------------------------------------------------------------------------------------------

%= = = = = = = = = = = = = = = = = = = = = = = = = = = = = = = = = = = = =
\subsection{Non-zero temperature}
\label{sec:therm}
The final factor considered here is how temperature affects the state. For a canonical ensemble in thermal equilibrium at temperature $T$, with Hamiltonian $H$, the density matrix is~\cite{barnett97},
\begin{eqnarray}
\rho(T)=\frac{\exp(- \frac{H}{k_B T})}{\mbox{Tr}\left[\exp(-\frac{H}{k_B T})\right]}
\end{eqnarray}
where $k_B$ is the Boltzman constant. Again, the system can be understood by considering the effective Hamiltonian given by Eq. (\ref{eq:Heff}). It can be shown, using Eq.~(\ref{eq:FQmix}), that the QFI is
\begin{eqnarray}
F_Q = N^2 \tanh^2 \left( \Delta E \over 2 k_B T \right).
\label{eq:QFI_TG}
\end{eqnarray}

Figure~\ref{fig:QvT} shows the QFI as a function of temperature for different interaction strengths. The thin (black) lines show the numerically calculated QFI. This was done using Eq.~(\ref{eq:FQmix}) and the lowest 20 energy levels were used to create the thermal density matrix. The density matrix was also created with larger numbers of energy levels to check the convergence, which showed that there was no significant change to the QFI. As expected, the QFI decreases as the temperature increases, because higher excited states become populated. However, the decrease in $F_Q$ is less rapid as the interaction strength is increased. The thick (blue) line in Fig.~\ref{fig:QvT} shows the analytic result for the Tonks-Girardeau case, which agrees well with the numerical results. For the NOON state the decrease is so rapid that $F_Q \approx 0$ on this scale. The red crosses in Fig.~\ref{fig:QvT} show the QFI at zero temperature. For interactions less than that required to make a NOON state, the energy gap between the ground and first excited state increases, so the system becomes less affected by temperature. However, the QFI at zero temperature is also reduced due to the loss of the binary superposition.

For a superposition in the Tonks-Girardeau regime, as the number of atoms is increased, the superposition becomes more robust to temperature, however, this is due to the increase in the energy level splitting between the ground and first excited state. As shown in Fig.~\ref{fig:DEvN}, the splitting becomes almost constant for more than ten atoms. This suggests the robustness to temperature will reach a constant value for all numbers of atoms and a small barrier, however this robustness can be improved by increasing the height of the barrier.
%= = = = = = = = = = = = = = = = = = = = = = = = = = = = = = = = = = = = = = = = =

%=============================================================
\section{Experimental realization}
In this section we consider the experimental realization of 100 ${}^7$Li atoms confined to a tight ring trapping potential of radius $50\mu$m. This gives a mean atom spacing of $3.1\mu$m. Here we choose a modest transverse oscillator frequency $\omega_{\perp} = 9$ kHz, which gives a transverse radial confinement of $a_{\perp} = 1 \mu$m. The superposition is most readily created for a large barrier height, however, this strongly couples states other than $|0\rangle$ and $|N\rangle$. As a compromise we choose half the limiting value for the energy gap in the Tonks-Girardeau regime, $\Delta E \approx 25 E_0 \approx 45 \hbar $ Hz.

The superposition is created by adiabatically increasing the angular velocity of the stirring barrier to $0.29 \times 2\pi$ Hz. This should be done with a barrier with a width less than $0.5\mu$m, according to Eq.~(\ref{eq:width}), so that the energy gap at the avoided crossing is not significantly effected by a finite width barrier. From Eqs.~(\ref{eq:lorentzdist}) and~(\ref{eq:gamma}) we see the accuracy of the angular velocity must be within $0.01 \times 2\pi$  Hz to achieve a well balanced superposition. From Eq.~(\ref{eq:domdg1}) we also see that the rate at which the rotation can be increased, with only a small probability of excitation, must be much less than $26\times 2\pi$ Hz s${}^{-1}$. If we take the extreme case that the atoms are in thermal equilibrium with the environment, then, from Eq.~(\ref{eq:QFI_TG}) we would have to reduce the temperature of the condensate to less than $0.2$nK. This is due to the small energy gap at the crossing. However, if we assume the population in the ground state remains the same as when the condensate is created for a stationary barrier, then higher temperatures should still produce the superposition. Finally, to create the NOON state, Eq.~(\ref{eq:dgdt}) shows that the interaction strength, $g$, must be reduced at a rate much less than $1.7\times 10^{-39}$ kg m${}^3$s${}^{-3}$ to avoid excitations, which is equivalent to reducing the scattering length at a rate much less than $0.0044$ \AA$(\textrm{ms})^{-1}$. We should note that this is the rate required at the value of $g\approx LE_0/(2N)$ where the NOON state is reached and excitations are most likely. Much faster rates will be possible for larger interaction strengths. 
%=============================================================

%=============================================================
\section{Conclusion}
We have studied a spectrum of multi-particle superpositions of total angular momentum to understand how to make them large and experimentally accessible. Compared to NOON states, the results presented here show that superpositions using strongly correlated atoms are far more robust to potential decoherence, finite temperature and allow a faster evolution of experimental parameters without a significant probability of excitations. Although it looks unlikely that a large NOON state can be created by changing the stirring rate of the barrier, a more realistic approach would be to create a superposition in the strongly correlated regime first and then slowly ramp down the interaction strength. This could be achieved via a Feshbach resonance. Experimentally realising this scheme will provide a means to probe the boundary between quantum and classical mechanics, and improve precision measurements beyond the shot noise limit.
%==============================================================

\section{Acknowledgments}
We acknowledge stimulating discussions with Jacob Dunningham and Thomas Ernst. This work was supported by the Marsden Fund (Contract No. MAU0706), administered by the Royal Society of New Zealand

%===============================================================

%================================================================

\end{document}